# MECHANISM OF FERROELECTRIC THIN FILMS SELF-POLARIZATION PHENOMENON.

M.D.Glinchuk[*], A.N.Morozovska[**]

[*]Institute for Problems of Materials Science, National Academy
of Science of Ukraine, Krjijanovskogo 3, 03142 Kiev, Ukraine, glin@materials.kiev.ua

[**]V. Lashkaryov Institute of Semiconductor Physics, NAS of Ukraine,
41, pr. Nauki, 03028 Kiev, Ukraine, morozo@mail.i.com.ua

**Abstract**

In present work we calculated the three components of polarization in phenomenological theory framework by consideration of three Euler-Lagrange equations, which include mismatch effect and influence of misfit dislocations, surface tension and depolarization field. These equations were solved with the help of variational method proposed by us earlier. This approach lead to the free energy in the form of algebraic equation of different powers of polarization components with the coefficients dependent on film thickness, mismatch effect, temperature and other parameters. Several new terms proportional to misfit strain appeared in the free energy expression: built-in electric field normal to the surface originated from piezoelectricity in vicinity of surface even in the cubic symmetry of bulk ferroelectrics, renormalization of bulk transition temperature via electrostriction, odd powers of normal to the surface component of polarization. The obtained free energy made it possible to calculate all properties of the film by conventional procedure of minimization. As an example we calculated phase diagrams of PZT 50/50 films on different substrates, that lead to compressive or tensile strain $u$ in coordinates $T$, $u$, $l$, where $l$ is a film thickness, for two cases − without and with external electric field, that compensate built-in field. It was shown that in both cases self-polarized phase does exist in definite regions of aforementioned coordinates, its nature being mismatch effect. The calculations of pyroelectric coefficient and dielectric permittivity dependencies on $T$, $u$ and $l$ had shown the unusual behaviour of these quantities, e.g. existence of pyroelectricity at $l < l_{cr}$, where $l_{cr}$ is critical thickness of ferroelectric-paraelectric phase transitions.

<u>Keywords</u>: surface tension, thin film, mismatch effect, phase transition.

PACS: 77.80.-e, 77.55.+f, 68.03.Cd, 68.35.Gy

## 1. INTRODUCTION

The self-polarization phenomenon in ferroelectric thin films, i.e. the presence of a significant amount (~90%) polarization in the normal to the surface direction without any poling treatment, has been reported by many authors (see [1]-[6]). The self-polarization is advantageous for piezoelectric and pyroelectric applications since the conventional poling process could be skipped altogether and so the production cost and time can be reduced and non-uniformity of the polarization could be avoided.

The characteristic feature of self-polarized film was shown to be the asymmetry of hysteresis loops relatively to the point of zero external electric field. Since this asymmetry looks like that in the bulk ferroelectrics under the application of external dc-field, it is widely believed that the ferroelectric thin films are self-polarized by a built-in electric field across the film. There were several suppositions about this field origin in PZT films, which are usually used for pyroelectric and piezoelectric applications. The authors of [3] suggested that it was an electric field of Schottky barrier between PZT film and Pt bottom electrode and thus it is localized in the immediate vicinity of the bottom electrode and decreased with the distance into the film. No self-polarization was hence observed in their PZT films thicker than 700nm. The space charge originated from oxygen vacancies, trapped electrons and asymmetry of their distribution was supposed to be the source of built-in electric field across the film [4], [5]. However there were no confirmations of these suppositions neither on the basis of calculations nor with the help of some special experimental investigations.

Recently we had shown [7], that mechanical tension originated from mismatch effect between a film and a substrate produces a built-in electric field in the film allowing for the absence of inversion centre in the vicinity of a film surface that results into piezoelectric effect existence. This field oriented in normal to the surfaces direction was shown to be inversely proportional to the film thickness so that it can be neglected for thick enough films. This result and asymmetry of calculated hysteresis loops are in a good agreement with aforementioned experimental data. Moreover some unusual for cubic symmetry odd power polarization terms related to the mismatch effect appeared in the free energy, along with the contribution of renormalized bulk transition temperature. Because of this we supposed in [7], that the film-substrate mismatch effect can be the origin of self-polarization phenomenon. However the calculations were performed only for $P_Z$ component of polarization (z axis is normal to the film surface). Since there is a coupling between in plane components of polarization $P_X$, $P_Y$ and $P_Z$ it was apriory unclear if nonzero in-plane polarization can appear at $P_Z \neq 0$ and so destroy the film self-polarization along z-axis. Because of this the calculations for the case of $P_X$, $P_Y$ and $P_Z$ components of polarization appeared to be necessary to carry out.

In the present work we performed calculations of the ferroelectric film properties allowing for mismatch effect in phenomenological theory framework for three components of polarization. The obtained dependences of the components on misfit strain $U_m$, temperature $T$ and film thickness $l$ allowed to calculate phase diagrams in the coordinates $(T, U_m)$, $(U_m, l)$ and $(T, l)$. For the parameters of PZT ferroelectric film the regions of self-polarized phase ($P_X = P_Y = 0$, $P_Z \neq 0$) existence was obtained. The renormalized free energy in the form of different powers of $P_X$, $P_Y$ and $P_Z$ with coefficients dependent on the film thickness and misfit strain was obtained. For the first time both the gradient of polarization and depolarization field influence (in the course of ref. [8] [9]) has been taken into account, in contrast to ref. [10], where permanent polarization along the film was considered. Misfit dislocations influence was considered in [11] within the model [12]. Our results open the way of all the properties calculations on the basis of conventional minimization of free energy with renormalized coefficients instead of differential Euler-Lagrange equations solutions. As an example we calculated the thickness dependence of dielectric permittivity and pyroelectric coefficient Π without external electric field and for the case when external electric field compensates the mismatch induced built-in field. In the letter case Π=0 at thickness $l$ smaller than the critical one, while in the former case Π≠0 for all values of $l$ because of mismatch induced self-polarization.

## 2. FREE ENERGY FUNCTIONAL

Let us consider ferroelectric thin film with the thickness $l$ $(-l/2 \leq z \leq l/2)$ and external electric field **E**. In phenomenological theory approach free energy functional can be written in the form:

$$\Delta G = G - G_0 = \int g_V dV + \int g_S ds. \tag{1a}$$

Here the first and the second integral reflect polarization dependent contribution of the bulk and the surface of the film, $G_0$ is polarization independent part of the free energy. The bulk free energy density $g_V$ can be represented as:

$$g_V = g_{0V} + \frac{1}{2}\left(\delta_Z\left(\frac{dP_Z}{dz}\right)^2 + \delta_X\left(\frac{dP_X}{dz}\right)^2 + \delta_X\left(\frac{dP_Y}{dz}\right)^2\right) - \mathbf{P}\,\mathbf{E} - P_Z \frac{E_d}{2} \tag{1b}$$

Here the energy $g_{0V}$ represents itself the expansion over polarization powers with coefficients renormalized by mechanical tension via electrostriction on the base of procedure developed in [10]; $E_d$ is depolarization field, its value for the case of single-domain insulator film with super-conducting electrodes can be written in the form [9]:

$$E_d = 4\pi\left(\overline{P}_Z - P_Z(z)\right). \tag{2}$$

Hereafter the bar over a letter denoting a physical quantity represents the spatial averaging over the film thickness, e.g. $\overline{P}_Z \equiv \frac{1}{l}\int_{-l/2}^{l/2} dz P_Z(z)$. Allowing for surface energy is related to surface tension [13], it is possible to represent the second term in Eq. (1) similarly to [7], [14]:

$$G_s = \sum_{i=1}^{2}\int \mu_i u_{xx}^{(i)} u_{yy}^{(i)} dxdy. \tag{3a}$$

Here parameters $\mu$ and $u_{jj}$ ($j = x, y$) are respectively the surface tension coefficient and strain tensor components, $i = 1, 2$ reflects the contribution of the film two surfaces.

In what follows we will consider two main contributions to the strain tensor, proposed in [7] related to surface polarization $P_Z(\pm l/2)$ via piezoelectric effect that exists even in a cubic symmetry lattice near the film surface, and the second is related to mismatch effects discussed in the introduction. Therefore

$$u_{xx}^{(i)} = u_{xxm}^{(i)} + d_{xxz}^{(i)} P_{zi}^{(i)}, \quad u_{yy}^{(i)} = u_{yym}^{(i)} + d_{yyz}^{(i)} P_{zi}^{(i)}, \tag{3b}$$

where $z_1 = l/2$, $z_2 = -l/2$, $d_{jjk}$ is the coefficient of piezoelectric effect, $u_{jjm}$ is the tensor of mechanical strain that is proportional to the difference of the lattice constants and thermal expansion coefficients of a substrate and a film as well as to the growth imperfections. In what follows we will consider an epitaxial film with bulk cubic symmetry, e.g. PbZr$_k$Ti$_{1-k}$O$_3$ (k/1-k PZT). In such case $u_{xx}^{(i)} = u_{yy}^{(i)}$, $d_{xxz}^{(i)} = d_{yyz}^{(i)} \equiv d_{13}$ and $u_{xxm}^{(i)} = u_{yym}^{(i)} \equiv u_m^{(i)}$ so the product $u_{xx}^{(i)} u_{yy}^{(i)}$ in Eq. (4) can be rewritten as $u_{xx}^{(i)} u_{yy}^{(i)} = \left(u_m^{(i)} + d_{13} P_{zi}^{(i)}\right)^2$, where the term $u_m^{(i)}$ is independent on polarization, while the other terms are defined the surface free energy. In what follows we will consider the realistic situation of the film on the substrate with free-standing upper surface, where parameters $u_m^{(1)} = 0$ and $u_m^{(2)} = U_m$. For the description of the second order phase transitions in the films with perovskite structure the free energy (1) with respect to the Eqs.(2-4) acquires the form $\Delta G = \Delta G_V + \Delta G_S$:

$$\Delta G_V = \frac{1}{l} \int_{-l/2}^{l/2} dz \left[ \begin{array}{l} \frac{\alpha_Z(T,U_m)}{2} P_Z^2(z) + \frac{\alpha_X(T,U_m)}{2}\left(P_X^2(z) + P_Y^2(z)\right) + \frac{\eta_X}{2} P_X^2(z) \cdot P_Y^2(z) + \\ + \frac{\eta_Z}{2} P_Z^2(z)\left(P_X^2(z) + P_Y^2(z)\right) + \frac{\beta_Z}{4} P_Z^4(z) + \frac{\beta_X}{4}\left(P_X^4(z) + P_Y^4(z)\right) + \\ + \frac{1}{2}\left(\delta_Z \left(\frac{dP_Z(z)}{dz}\right)^2 + \delta_X \left(\frac{dP_X(z)}{dz}\right)^2 + \delta_X \left(\frac{dP_Y(z)}{dz}\right)^2\right) - P_Z(z)\left(E_Z + 2\pi\left(\overline{P}_Z - P_Z(z)\right)\right) - \\ - P_X(z) E_X - P_Y(z) E_Y \end{array} \right]$$

$$\Delta G_S = \frac{\delta_Z}{2l}\left[\frac{P_Z^2(l/2)}{\lambda_{Z1}} + \frac{\left(P_Z(-l/2) + P_m\right)^2}{\lambda_{Z2}}\right] + \frac{\delta_X}{2\lambda_X l}\left[P_X^2(l/2) + P_X^2(-l/2) + P_Y^2(l/2) + P_Y^2(-l/2)\right]$$

(4)

Hereinafter $\beta_{X,Z} > 0$, $\delta_{X,Z} > 0$ and we introduced the extrapolation lengths $\lambda_X = \lambda_Y$, $\lambda_{Z1,2}$ and "misfit-induced" surface polarization $P_m$ in the form:

$$\lambda_{Z1,2} = \frac{\delta_Z}{2\mu_{1,2} d_{13}^2}, \qquad P_m = \frac{U_m}{d_{13}}. \qquad (5)$$

Since the signs of the parameters $U_m$ and $d_{13}$ can be positive or negative both $P_m > 0$ and $P_m < 0$ are expected. The extrapolation lengths $\lambda_{X,Z1,2}$ can be only positive because $\mu_{1,2} > 0$. The renormalized by mechanical tension coefficient $\alpha$ in (4) has the form [10]:

$$\alpha_{X,Z}(T) = \alpha_T\left(T - T_C^{X,Z}\right), \quad T_C^Z = T_C + \frac{2Q_{12} U_m^*}{\alpha_T(S_{11} + S_{12})}, \quad T_C^X = T_C + \frac{(Q_{11} + Q_{12}) U_m^*}{\alpha_T(S_{11} + S_{12})}. \qquad (6)$$

Here parameters $T_C$, $\alpha_T$, $Q_{11}$, $Q_{12}$ and $S_{11}$, $S_{12}$ are respectively ferroelectric transition temperature, inverse Curie constant, electrostriction coefficient and elastic modulus regarded known for the bulk material. In (6) we take into account, that in accordance with [11] the "effective" substrate constant is renormalized by misfit dislocations appeared at critical thickness $l_d$, namely:

$$U_m^*(T,l) = \frac{b^*(T,l) - a(T)}{b^*(T,l)}, \qquad b^*(T,l) = \begin{cases} b(T)\left(1 - U_m(T)\left(1 - \frac{l_d}{l}\right)\right), & l > l_d \\ b(T), & l \leq l_d \end{cases} \qquad (7)$$

Temperature dependence in Eq.(7) is related to thermal expansion coefficients of the film ($\rho_a$) and substrate ($\rho_b$) so that $a(T) \approx a(T_g)(1+(T-T_g)\rho_a)$ and $b(T) \approx b(T_g)(1+(T-T_g)\rho_b)$ are film material and substrate lattice constants respectively, $T_g$ is film growth temperature. However on the surface $z = -l/2$ (i.e. in the surface energy $\Delta G_S$ and thus in (5)) the real misfit strain $U_m(T)$ exists, namely:

$$U_m(T) = \frac{b(T)-a(T)}{b(T)} \approx \frac{b(T_g)-a(T_g)}{b(T_g)} + (\rho_b - \rho_a)(T-T_g) \tag{8}$$

Having substituted (8) into (7) one obtains:

$$U_m^*(T,l) = \frac{b^*(T,l)-a(T)}{b^*(T,l)} = \begin{cases} \dfrac{l_d}{l}\dfrac{U_m(T)}{(1-U_m(T)l_d/l)}, & l > l_d \\ U_m(T), & l \leq l_d \end{cases} \tag{9}$$

## 3. FREE ENERGY WITH RENORMALIZED COEFFICIENTS

The coupled equations for the polarization components can be obtained by variation over polarization of free energy functional (4). This yields the following Euler-Lagrange equations with the boundary conditions:

$$\begin{cases} P_Z\left(\alpha_Z + \eta_Z(P_X^2+P_Y^2)\right) + \beta_Z P_Z^3 - \delta_Z\left(\dfrac{d^2 P_Z}{dz^2}\right) = E_Z + 4\pi(\bar{P}_Z - P_Z), \\ \left(P_Z + \lambda_{Z1}\dfrac{dP_Z}{dz}\right)\bigg|_{z=l/2} = 0, \quad \left(P_Z - \lambda_{Z2}\dfrac{dP_Z}{dz}\right)\bigg|_{z=-l/2} = -P_m. \end{cases} \tag{10a}$$

$$\begin{cases} P_X\left(\alpha_X + \eta_Z P_Z^2 + \eta_X P_Y^2\right) + \beta_X P_X^3 - \delta_X\left(\dfrac{d^2 P_X}{dz^2}\right) = E_X, \\ \left(P_X + \lambda_X\dfrac{dP_X}{dz}\right)\bigg|_{z=l/2} = 0, \quad \left(P_X - \lambda_X\dfrac{dP_X}{dz}\right)\bigg|_{z=-l/2} = 0. \end{cases} \tag{10b}$$

$$\begin{cases} P_Y\left(\alpha_X + \eta_Z P_Z^2 + \eta_X P_X^2\right) + \beta_X P_Y^3 - \delta_X\left(\dfrac{d^2 P_Y}{dz^2}\right) = E_Y, \\ \left(P_Y + \lambda_X\dfrac{dP_Y}{dz}\right)\bigg|_{z=l/2} = 0, \quad \left(P_Y - \lambda_X\dfrac{dP_Y}{dz}\right)\bigg|_{z=-l/2} = 0. \end{cases} \tag{10c}$$

Let us find as it was proposed earlier in [15], [16] the approximate solution of the nonlinear Eqs.(10) by the direct variational method. We will choose the one-parametric trial functions similarly to [7], in the form of solutions of linearized Eqs.(10) with external field components $E_Z$, $E_X$, $E_Y$, that satisfy the boundary conditions. In the case of different extrapolation lengths $\lambda_{Z1} \neq \lambda_{Z2}$ trial functions become very cumbersome (see e.g. [17]). In order to demonstrate the mismatch effect influence more clearly, hereinafter we put $\lambda_{Z1} = \lambda_{Z2} = \lambda_Z$ and use the following trial functions:

$$P_Z(z) = P_{VZ}[1-\varphi(z)] - \frac{P_m}{2}[\varphi(z) - \xi(z)], \qquad P_{X,Y}(z) = P_{V\,X,Y}[1-\phi(z)]. \tag{11}$$

The variational parameters - amplitudes $P_{VX,Y,Z}$ must be determined by the minimisation of the free energy (4). Hereinafter we used the following functions:

$$\varphi(z) = \frac{ch(z/l_Z)}{ch(l/2l_Z) + (\lambda_Z/l_Z)sh(l/2l_Z)}, \qquad \xi(z) = \frac{sh(z/l_Z)}{sh(l/2l_Z) + (\lambda_Z/l_Z)ch(l/2l_Z)}, \tag{12a}$$

$$\phi(z) = \begin{cases} \dfrac{ch(z/l_X)}{ch(l/2l_X) + (\lambda_X/l_X)sh(l/2l_X)}, & \alpha_X > 0 \\ \dfrac{\cos(z/l_X)}{\cos(l/2l_X) + (\lambda_X/l_X)\sin(l/2l_X)}, & \alpha_X < 0 \end{cases} \tag{12b}$$

Here $l_Z$ and $l_X$ are respectively the longitudinal and transverse characteristic lengths: $l_Z = \sqrt{\delta_Z/(4\pi + \alpha_Z)} \approx \sqrt{\delta_Z/4\pi}$ and $l_X = \sqrt{\delta_X/|\alpha_X|}$. It follows from Eq.(6), that sign of $\alpha_X$ strongly depends on $U^*_m$ sign and $T_C$ value. For the most of ferroelectrics $l_Z \sim 1 \div 10 \,\overset{\circ}{A}$ [18], and so hereinafter we assume that: $l \gg l_Z$, $l_X \gg l_Z$, $l \le l_X$, $\lambda_Z \gg l_Z$, $\lambda_X \sim l_X$.

Hereinafter we supposed that $E_{X,Y} = 0$, $E_Z = E_0$. Integration in the expression (4) with the trial functions (11)-(12) leads to the following form of the free energy with renormalized coefficients:

$$\Delta G(P_V) = \begin{bmatrix} \dfrac{A_m}{2}P_{VZ}^2 + \dfrac{D_m}{3}P_{VZ}^3 + \dfrac{B_m}{4}P_{VZ}^4 + \dfrac{A_X}{2}\left(P_{VX}^2 + P_{VY}^2\right) + \dfrac{B_X}{4}\left(P_{VX}^4 + P_{VY}^4\right) + \\ + \dfrac{C_{XY}}{2}P_{VX}^2 P_{VY}^2 + \dfrac{F_{XZ}}{2}P_{VZ}^2\left(P_{VX}^2 + P_{VY}^2\right) - K_m P_{VZ}\left(P_{VX}^2 + P_{VY}^2\right) - P_{VZ}(E_0 - E_m) \end{bmatrix}. \tag{13}$$

The renormalized coefficients in (13) have the following form:

$$A_m = (\alpha_Z + 4\pi\overline{\varphi(z)})(1 - \overline{\varphi(z)}) + \frac{3\beta_Z}{4}P_m^2 \overline{(1-\varphi(z))^2(\varphi^2(z) + \xi^2(z))} \approx$$
$$\approx \alpha_T\left(T - T_C - \frac{2Q_{12}U^*_m}{\alpha_T(S_{11} + S_{12})}\right)\left(1 - \frac{1}{h(1+\Lambda_Z)}\right) + \frac{4\pi}{h(1+\Lambda_Z)} + \frac{\beta_Z}{d_{13}^2}\frac{3U_m^2}{4h(1+\Lambda_Z)^2} \tag{14}$$

$$B_m = \beta_Z \overline{(1-\varphi(z))^4} \approx \beta_Z\left(1 - \frac{3}{h(1+\Lambda_Z)}\right), \tag{15}$$

$$D_m = -\frac{3\beta_Z}{2}P_m \overline{(1-\varphi(z))^3 \varphi(z)} \approx -\frac{3\beta_Z}{2d_{13}}\frac{U_m}{h(1+\Lambda_Z)}, \tag{16}$$

$$E_m = 2\pi\overline{\varphi(z)(1-\varphi(z))}P_m - \frac{\beta_Z}{8}P_m^3 \overline{(\varphi^3(z) + 3\varphi(z)\xi^2(z))(1-\varphi(z))} \approx$$
$$\approx \frac{2\pi}{d_{13}}\frac{U_m}{h(1+\Lambda_Z)} \tag{17}$$

$$A_X = \alpha_X \overline{(1-\phi(z))} + \frac{\eta_Z}{4} P_m^2 \overline{(1-\phi(z))^2 (\varphi^2(z)+\xi^2(z))} \approx$$
$$\approx \alpha_T \left(T - T_C - \frac{(Q_{11}+Q_{12})U_m^*}{\alpha_T(S_{11}+S_{12})}\right)\frac{1}{(1+R/h)} + \frac{\eta_Z}{d_{13}^2}\frac{U_m^2}{4h(1+\Lambda_Z)^2}\frac{1}{(1+R/h)^2} \quad (18)$$

$$B_X = \beta_X \overline{(1-\phi(z))^4} \approx \frac{\beta_X}{(1+R/h)^4}, \quad C_{XY} = \eta_X \overline{(1-\phi(z))^4} \approx \frac{\eta_X}{(1+R/h)^4} \quad (19)$$

$$F_{XY} = \eta_Z \overline{(1-\phi(z))^2(1-\varphi(z))^2} \approx \frac{\eta_Z}{(1+R/h)^2}\left(1-\frac{2}{h(1+\Lambda_Z)}\right), \quad (20)$$

$$K_m = \eta_Z P_m \overline{(1-\phi(z))^2(1-\varphi(z))\varphi(z)} \approx \frac{\eta_Z}{d_{13}}\frac{U_m}{h(1+\Lambda_Z)(1+R/h)^2}. \quad (21)$$

Here the following dimensionless parameters are introduced $h = \frac{l}{2\sqrt{\delta_Z/4\pi}}$, $\Lambda_Z = \frac{\lambda_Z}{\sqrt{\delta_Z/4\pi}}$, $R = \frac{\delta_X/|\alpha_X|}{\lambda_X\sqrt{\delta_Z/4\pi}}$. The approximate expressions for the renormalized coefficients (14)-(21) are valid at $h \gg 1$, $\Lambda_Z \gg 1$ and $l \leq \delta_X/|\alpha_X|$ (see Appendix A). Note, that odd power $P_{VZ}^3$, $P_{VZ}(P_{VX}^2+P_{VY}^2)$ in Eq.(13) is unusual for cubic symmetry perovskite structure ferroelectrics, these terms as well as $E_m$ are absent at $P_m = 0$ (see Eqs.(16), (17), (21)), i.e. they are related to mismatch effect. In accordance to Eq.(9): $U_m^*(T,h) = U_m(T)$ at $h \leq h_d$ and $U_m^*(T,h) = \frac{U_m(T)h_d}{h(1-U_m(T)(1-h_d/h))}$ at $h > h_d$.

## 4. PHASE DIAGRAMS WITH SELF-POLARIZED REGIONS

The coupled equations for the amplitudes $P_V$ can be obtained by variation of the renormalized free energy (13). This yields the following coupled equations:

$$\begin{cases} [A_m + F_{XZ}(P_{VX}^2 + P_{VY}^2)]P_{VZ} + B_m P_{VZ}^3 + D_m P_{VZ}^2 - K_m(P_{VX}^2 + P_{VY}^2) = E_0 - E_m, \\ [A_X + C_{XY}P_{VY}^2 + F_{XZ}P_{VZ}^2 - 2K_m P_{VZ}]P_{VX} + B_X P_{VX}^3 = 0, \\ [A_X + C_{XY}P_{VX}^2 + F_{XZ}P_{VZ}^2 - 2K_m P_{VZ}]P_{VY} + B_X P_{VY}^3 = 0. \end{cases} \quad (22)$$

The last two equations for $P_{VX,Y}$ can be solved independently and their solutions $P_{VX,Y}(P_{VZ})$ must be substituted into the first equation for $P_{VZ}$ as well as into the Eq.(13). Similarly to the bulk ferroelectrics these equations allowed to analyse possible equilibrium orientations of polarization for some values of the coefficients and so for some film thickness, temperature and misfit strain regions. In this way we obtained four different kind of phases, namely:

a). "Asymmetrical" $ac$-phase ($P_X \neq 0 \leftrightarrow P_Y = 0$, $P_Z \neq 0$) and $a$-phase ($P_X \neq 0$, $P_{Y,Z} = 0$):

$$P_{VX} \equiv 0, \quad P_{VY}^2 = -\frac{A_X + F_{XZ}P_{VZ}^2 - 2K_m P_{VZ}}{B_X}, \quad G_A(P_{VZ}) = \min, \tag{23a}$$

$$G_A(P_{VZ}) = \begin{bmatrix} \left(A_m - \frac{F_{XZ}A_X - 2K_m^2}{B_X}\right)\frac{P_{VZ}^2}{2} + \left(D_m + \frac{3K_m F_{XZ}}{B_X}\right)\frac{P_{VZ}^3}{3} + \\ + \left(B_m - \frac{F_{XZ}^2}{B_X}\right)\frac{P_{VZ}^4}{4} - P_{VZ}(E_0 - E_m) - \frac{A_X^2}{4B_X} \end{bmatrix}. \tag{23b}$$

b). "Symmetrical" $r$-phase ($P_X = P_Y \neq 0, \quad P_Z \neq 0$) and $aa$- phase ($P_X = P_Y \neq 0, \quad P_Z = 0$):

$$P_{VX}^2 = P_{VY}^2 = -\frac{A_X + F_{XZ}P_{VZ}^2 - 2K_m P_{VZ}}{B_X + C_{XY}}, \quad G_S(P_{VZ}) = \min, \tag{24a}$$

$$G_S(P_{VZ}) = \begin{bmatrix} \left(A_m - \frac{F_{XZ}A_X - 2K_m^2}{B_X + C_{XY}}\right)\frac{P_{VZ}^2}{2} + \left(D_m + \frac{3K_m F_{XZ}}{B_X + C_{XY}}\right)\frac{P_{VZ}^3}{3} + \\ + \left(B_m - \frac{F_{XZ}^2}{B_X + C_{XY}}\right)\frac{P_{VZ}^4}{4} - P_{VZ}(E_0 - E_m) - \frac{A_X^2}{2(B_X + C_{XY})} \end{bmatrix}. \tag{24b}$$

c). "Unipolar" $c$-phase ($P_X = P_Y = 0, \quad P_Z \neq 0$):

$$P_{VX} = P_{VY} = 0, \quad P_{VZ} \neq 0, \quad G_C(P_{VZ}) = \min, \tag{25a}$$

$$G_C(P_{VZ}) = \left[A_m \frac{P_{VZ}^2}{2} + D_m \frac{P_{VZ}^3}{3} + B_m \frac{P_{VZ}^4}{4} - P_{VZ}(E_0 - E_m)\right]. \tag{25b}$$

d). Paraelectric $pe$-phase ($P_X = P_Y = P_Z = 0$):

$$\min[G_A(P_{VZ})] \geq 0 \,\&\, \min[G_S(P_{VZ})] \geq 0 \,\&\, \min[G_C(P_{VZ})] \geq 0. \tag{26}$$

The condition of $c$-phase absolute stability

$$\{\min[G_C(P_{VZ})] < 0 \,\&\, \min[G_C(P_{VZ})] < \min[G_{S,A}(P_{VZ})]\} \tag{27}$$

determines the region of parameters $(U_m, T, h)$ where the unipolar phase of the film $\{P_{X,Y} \equiv 0, \quad P_Z \neq 0\}$ is energetically preferable. Many bulk ferroelectric materials are not unipolar. In general case thin films prepared from them need pre-poling. Could the unipolar state, i.e. self-polarized state, be induced by mismatch effect in a thin film? Induced by mismatch effect surface polarization $P_m|\,|z$ leads to the appearance of built-in field $E_m$ (see Eq.(17)) and the terms proportional to the odd powers of $P_z$ (see Eqs.(13), (16), (21)). All these factors tend to induce self-polarization in the film, and so it can be expected for some film-substrate pairs. Necessary conditions (film thickness $h$, misfit strain $U_m$, temperature $T$) are determined by material coefficients and can be calculated for each definite material on the basis of Eqs.(25), (27). Below we give the concrete example.

The widely used 50/50 PZT bulk samples have perovskite structure with all nonzero components $P_{X,Y,Z}$ (see case b)), i.e. they are not unipolar in the ferroelectric phase. Our calculations proved that the unipolar state in 50/50 PZT thin films could be induced by mismatch effect. Phase diagrams for 50/50 PZT films are presented below. We use material coefficients $S_{11} = 13.8 \cdot 10^{-12} \, m^2/H$, $S_{12} = -4.07 \cdot 10^{-12} \, m^2/H$ and $d_{13} = 47 \cdot 10^{-3} V \cdot m/H$ [19], all other material coefficients are from [20]. Note, that for PZT films on typical metal-oxide substrates $|T - T_R| \leq 500 K$, $|\rho_b - \rho_a| \leq 5 \cdot 10^{-6}$ and $|(\rho_b - \rho_a)(T - T_R)| \leq 2.5 \cdot 10^{-3}$, whereas $\frac{b(T_R) - a(T_R)}{b(T_R)} \geq 1.5 \cdot 10^{-2}$. So real misfit strain in (9) appeared practically independent on temperature and thus one could use approximation $U_m \approx \frac{b(T_R) - a(T_R)}{b(T_R)}$.

To analyze the contributions of $E_m$ and of the terms with $P_Z$ odd power we performed the calculations for two cases: zero external field $E_0 = 0$ and compensating external field $E_0 = E_m$. In the first case we have complete contribution of mismatch effect, while in the second one $E_m$ is compensated by external field. The results of phase diagram calculations in coordinates $(T, U_m)$, $(U_m, h)$ and $(T, h)$ are represented in Figs.1-4 for $E_0 = 0$ (plots a) and $E_0 = E_m$ (plots b).

One can see the existence of self-polarized phase $\{P_{X,Y} \equiv 0, \; P_Z \neq 0\}$ in all the phase diagrams for broad region of compressive strain $U_m < 0$ and for some region of tensile strain $U_m > 0$ in the case $E_0 = 0$ only (see Figs.1,2). It is seen from the comparison of the phase diagrams depicted in *a* and *b* plots, that the region of self-polarized phase and *r*-phase existence are much larger at $E_0 = 0$ than at $E_0 = E_m$. It is because in the latter case a part of these two phases transforms respectively into paraelectric phase *pe* and *aa*-phase for $h > h_{cr}$ (see Fig.1). For the film with $h < h_{cr}$ at $E_0 = E_m$ (see Fig.2) *r*-phase disappeared completely, being transformed into *aa*-phase, although self-polarization conserved for some large enough $|U_m|$ value, while *pe* phase appears in the wide $(T, U_m)$ region. It should be underlined, that critical thickness depends on $U_m$ value (see Fig.5), so one must be careful considering the cases $h > h_{cr}$ and $h < h_{cr}$. At $E_0 = 0$ and $U_m = 0$ *r* and $c_p$ phase transforms into *aa* and *pe* phase (see dash-dotted line in Figs.1a,2a). It is obvious, that there are two parts of self-polarized phase, one induced mainly by $E_m$, that transforms into *pe* phase at $E_0 = E_m$ and another, that conserves at $E_0 = E_m$ would be related mainly to the terms with odd powers of $P_Z$ and renormalization of $T_c$. Because of this in the Figs. we introduced two parts of self-polarized phase denoting them as $c$ and $c_p$. Keeping in mind that characteristic feature of experimentally observed self-polarized phase was shown to be asymmetrical hysteresis loops, we calculated the loops for $c$ and $c_p$ phases by the way described in [7]. The results are depicted in Fig.6. One can see that the asymmetrical loop was obtained for $c$ phase only, while for $c_p$ phase the behaviour of polarization looks like that in paraelectric phase under electric field $E_m$. Because of this $c$ phase which exists at $E_0 = 0$ and $E_0 = E_m$ can be considered as true self-polarized phase.

The same type of different behaviour at $E_0 = 0$ and $E_0 = E_m$ was obtained for phase diagrams in ($U_m$, $h$) and ($T$, $h$) coordinates (see Figs.3-4). Some peculiarities appeared at $h = h_d$, i.e. at critical thickness of misfit dislocations appearance, because $U_m^*(h)$ decreases at $h > h_d$ (see Eq.(9)). The increase of $U_m$ value influence phase diagrams in ($U_m$, $h$) and ($T$, $h$) coordinates.

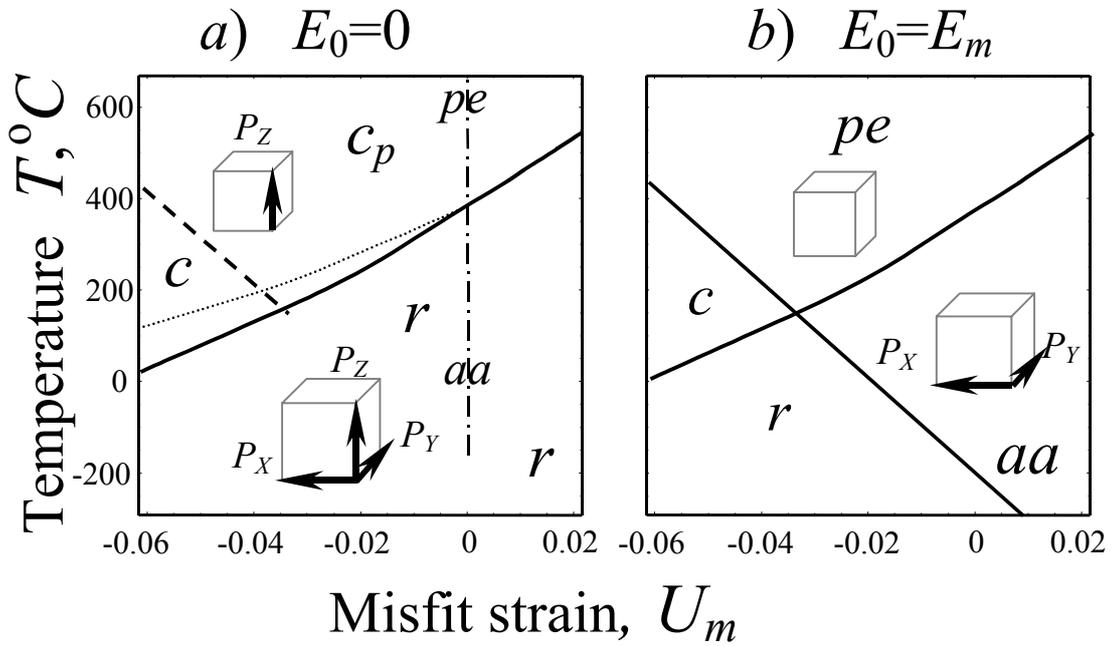

**Figure 1.** The phase diagram $T(U_m)$ for 50/50 PZT film for the following parameters: thickness $h = 60$, $h_d = 10$, $\Lambda = 10$, $R = 10$ and $E_0 = 0$ (**plot a**), $E_0 = E_m$ (**plot b**). Hereinafter dotted line inside $c$ phase designates the region with $0 \leq |P_{X,Y}/P_Z| \leq 0.01$.

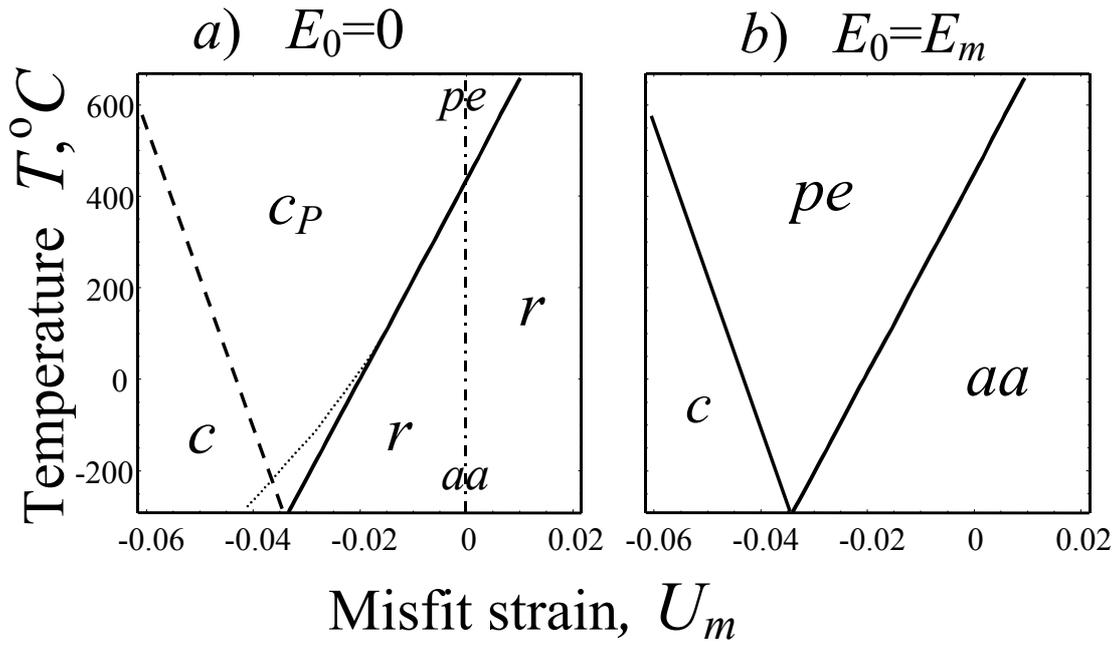

**Figure 2.** The phase diagram $T(U_m)$ for 50/50 PZT film for the following parameters: thickness $h = 20$, $h_d = 10$, $\Lambda = 10$, $R = 10$ and $E_0 = 0$ (**plot a**), $E_0 = E_m$ (**plot b**).

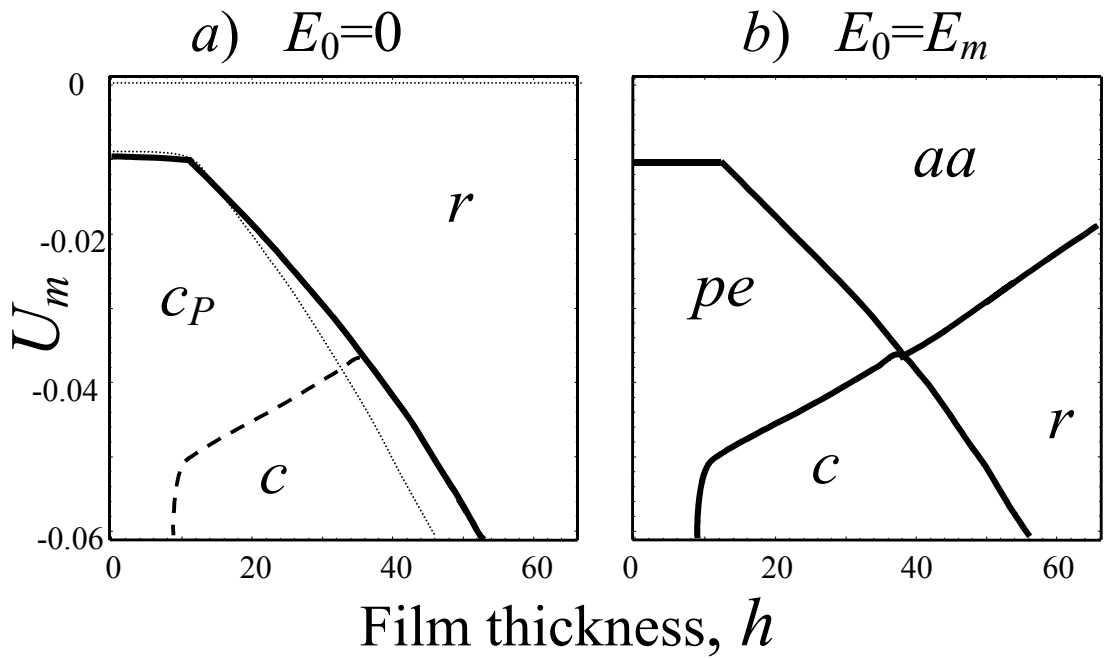

**Figure 3.** The phase diagram $Um(h)$ for 50/50 PZT film for the following parameters: T=25°C, $h_d = 10$, $\Lambda = 10$, $R = 10$ and $E_0 = 0$ (**plot a**), $E_0 = E_m$ (**plot b**).

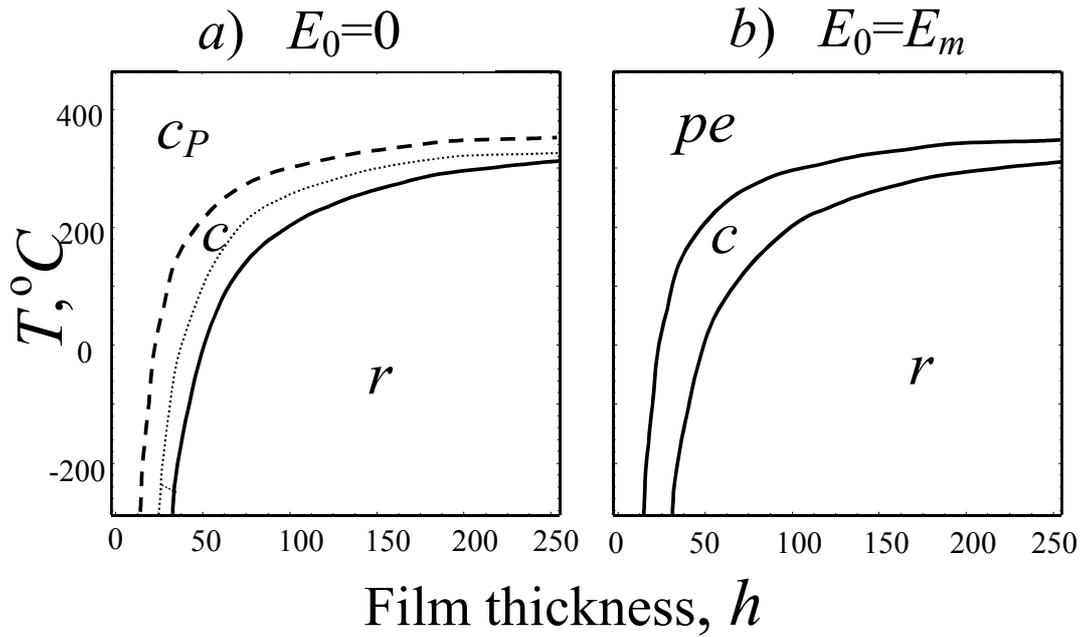

**Figure 4.** The phase diagram $T(h)$ for 50/50 PZT film for the following parameters: $U_m = -0.045$, $h_d = 10$, $\Lambda = 10$, $R = 10$ and $E_0 = 0$ (**plot a**), $E_0 = E_m$ (**plot b**).

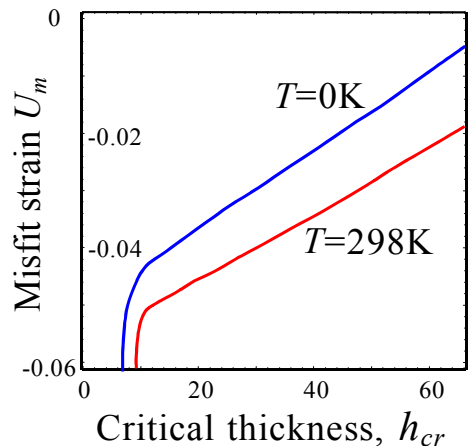

**Figure 5.** 50/50 PZT film critical thickness $h_{cr}$ over misfit strain $U_m$ at different temperatures $T$ for the following parameters $h_d = 10$, $\Lambda = 10$, $R = 10$, $E_0 = E_m$.

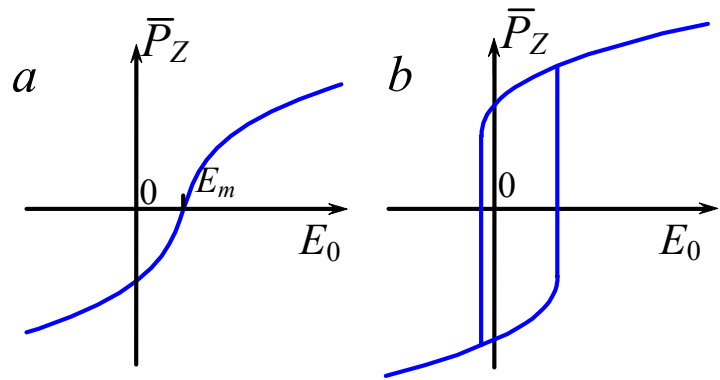

**Figure 6.** The typical hysteresis loops obtained at $U_m = -0.05$, T=25°C, $h_d = 10$, $\Lambda = 10$, $R = 10$ in phase $c_P$ ($h$=15, **plot a**) and in phase $c$ ($h$=30, **plot b**).

## 5. PROPERTIES OF THE FILMS

Free energy (13) with the renormalized coefficients (14)-(21) made it possible to calculate the properties as the functions of a film thickness, temperature and type of substrate, that defined $U_m$ value. As an example we carried out the calculations of pyroelectric coefficient and dielectric permittivity, allowing for these physical quantities are very sensitive to the existence of self-polarization.

### 5.1 Pyroelectric coefficient

Let us consider firstly the profile of the pyroelectric coefficient z-component $\Pi(z,T) = [1-\varphi(z)] \cdot dP_Z(z,T)/dT$, which determines the measurable pyroelectric current $j_z$. Coefficient $\Pi(z,T)$ exists in ferroelectric $c$-, $c_p$- and $r$-phases.

a). $\Pi(z,T)$ in $c$- and $c_p$ -phases. In our approximations functions $\varphi(z)$, $\xi(z)$ and coefficients $B_m$, $D_m$, $E_m$ are almost independent on temperature and so only $dA_m(T)/dT = \alpha_T$ will contribute to $dP_Z(z,T)/dT$. As it follows from (25), $\Pi(z,T)$ could be found from the system of equations:

$$\begin{cases} \Pi(z,T) \approx -\alpha_T \dfrac{P_{VZ}[1-\varphi(z)]}{A_m + 2D_m P_{VZ} + 3B_m P_{VZ}^2} \\ A_m P_{VZ} + D_m P_{VZ}^2 + B_m P_{VZ}^3 = E_0 - E_m \end{cases} \quad (28)$$

b). $\Pi(z,T)$ in $r$-phase. In our approximation functions $\varphi(z)$, $\xi(z)$ and coefficients $B_m$, $B_X$, $F_{XZ}$, $C_{XY}$, $D_m$, $E_m$ are almost independent on temperature and so only $dA_m(T)/dT = \alpha_T$ and $dA_X(T)/dT = \alpha_T/(1+R/h)$. As it follows from (24), $\Pi(z,T)$ could be found from the system of equations:

$$\begin{cases} \Pi(z,T) \approx \dfrac{-\alpha_T\left(1 - \dfrac{F_{XZ}}{(B_X + C_{XY})(1 + R/h)}\right) P_{VZ}[1-\varphi(z)]}{\left(A_m - \dfrac{F_{XZ}A_X - 2K_m^2}{B_X + C_{XY}}\right) + 2\left(D_m + \dfrac{3K_m F_{XZ}}{B_X + C_{XY}}\right) P_{VZ} + 3\left(B_m - \dfrac{F_{XZ}^2}{B_X + C_{XY}}\right) P_{VZ}^2} \\ \left(A_m - \dfrac{F_{XZ}A_X - 2K_m^2}{B_X + C_{XY}}\right) P_{VZ} + \left(D_m + \dfrac{3K_m F_{XZ}}{B_X + C_{XY}}\right) P_{VZ}^2 + \left(B_m - \dfrac{F_{XZ}^2}{B_X + C_{XY}}\right) P_{VZ}^3 = E_0 - E_m \end{cases} \quad (29)$$

One can see that the pyroelectric coefficient profiles (28)-(29) are always symmetrical, in contrast to the asymmetrical distribution of $P_Z(z,T)$ given by Eq.(11) with respect to Eqs.(12). It is clear from (28) that $\Pi(z,T) \neq 0$ in $c$-and $c_p$-regions (due to $P_{VZ}(T) \neq 0$). The average value of coefficient $\Pi$ one can obtain from (28), (29) and (12a). In the assumptions $h \gg 1$, $\Lambda_Z \gg 1$ and $l \leq \delta_X/|\alpha_X|$ we obtain that $1 - \varphi(z) \approx (1 - 1/h(1 + \Lambda_Z))$. This gives for $c$-, $c_p$-phase:

$$\overline{\Pi}(h,T) \approx -\alpha_T \frac{P_{VZ}}{A_m + 2D_m P_{VZ} + 3B_m P_{VZ}^2}\left(1 - \frac{1}{h(1+\Lambda_Z)}\right) \quad (30)$$

Thickness and temperature dependencies of this coefficient are depicted in Figs. 7 and 8 respectively for PZT (50/50) film on different substrates. The misfit strains for the substrates being summarised in Table1. One can see that $\overline{\Pi}(h,T)$ behaviour are strongly different for $E_0 = 0$ and $E_0 = E_m$ cases. The finite discontinuous changes on the curves in the Figs. 7-8 correspond to the boundaries between $c$ and $r$-phases (see thin dashed marks). The infinite discontinuous maximums at $E_0 = E_m$ correspond to the ferroelectric thickness-induced phase transition, that this transition is absent at $E_0 = 0$ due to the built-in field $E_m$ induced by mismatch effect. Therefore pyroelectricity exists in the broad region of film thickness and temperature and the values of the coefficient $\overline{\Pi}(h,T)$ are in good agreement with experimental values, which lays in the region $(20\div 200) nC/cm^2 K$ at room temperature $T = 25^0 C$ [11], [20].

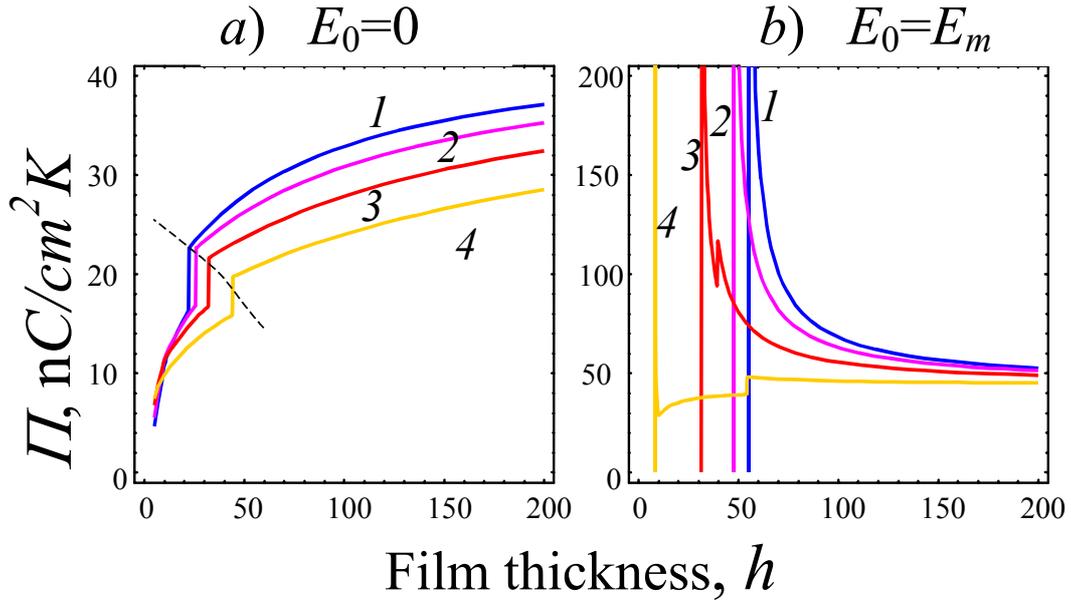

**Figure 7.** Pyroelectric coefficient $\overline{\Pi}(h,T)$ over film thickness $h$ for 50/50 PZT at $h_d = 10$, $T = 25^0 C$, $\Lambda = 10$, $R = 10$ and different substrates: Pt ($U_m = -0.024$, curve 1), SrTiO$_3$ ($U_m = -0.029$, curve 2), LSAT ($U_m = -0.039$, curve 3), LaAlO$_3$ ($U_m = -0.06$, curve 4).

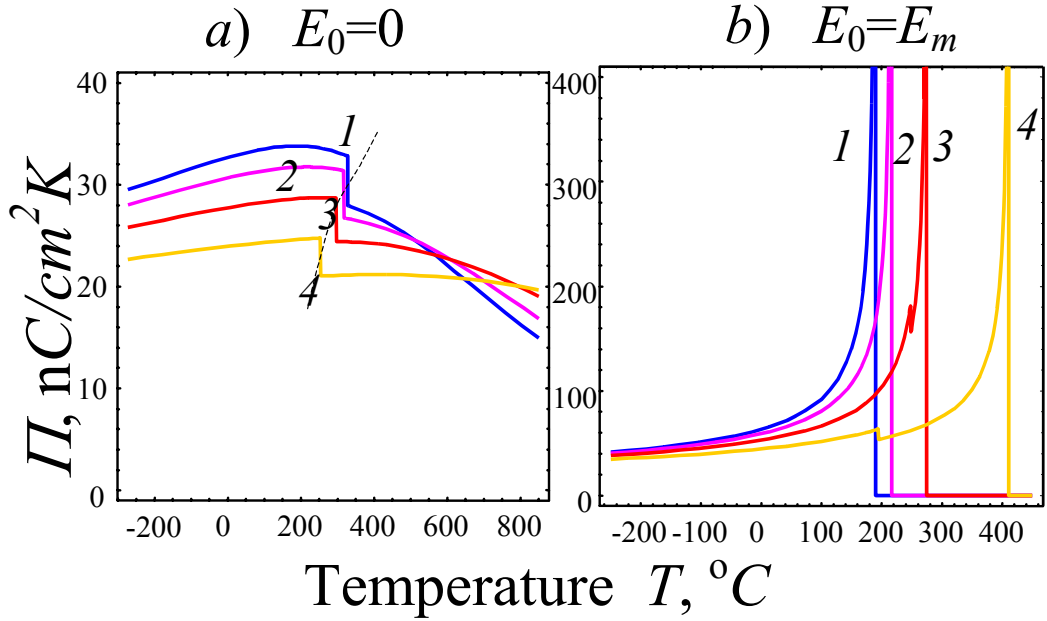

**Figure 8.** Pyroelectric coefficient $\overline{\Pi}(h,T)$ over temperature $T$ for 50/50 PZT film with thickness $h=100$ at $h_d=10$, $\Lambda=10$, $R=10$ and different substrates: Pt ($U_m=-0.024$, curve 1), SrTiO$_3$ ($U_m=-0.029$, curve 2), LSAT ($U_m=-0.039$, curve 3), LaAlO$_3$ ($U_m=-0.06$, curve 4).

**Table 1**. Misfit strain for 50/50 PZT film ($a$=4.0155A$^\circ$) on different substrates at $T=25^0 C$.

| Subs-trate | Subst. lattice constant. $b$, A$^\circ$ | Critical thickness of dislocations appearance $l_d$, A$^\circ$ [11], [12] | Misfit strain $U_m=(b-a)/b$ at $l<l_d$ | Effective substrate constant $b^*\approx b(1-U_m(1-l_d/l))$, A$^\circ$. at $l_d/l$=0.1;  0.25 | Effect. misfit strain $U^*_m=(b^*-a)/b^*$ at $l_d/l$=0.1;  0.25 |
|---|---|---|---|---|---|
| MgO | 4.211 | 29.1 | 0.046 | 4.037;  4.066 | 0.005;  0.012 |
| Au | 4.078 | 18.5 | 0.015 | 4.023;  4.032 | 0.002;  0.004 |
| Pt | 3.923 | 140 | -0.024 | 4.008;  3.994 | -0.002;  -0.005 |
| SrTiO$_3$ | 3.904 | 44.5 | -0.029 | 4.006;  3.989 | -0.002;  -0.007 |
| LSAT | 3.865 | 29.2 | -0.039 | 4.001;  3.978 | -0.004;  -0.009 |
| LaAlO$_3$ | 3.787 | 14.0 | -0.060 | 3.991;  3.957 | -0.006;  -0.015 |

### 5.2. Static dielectric permittivity

In accordance with definition [18], the average static linear dielectric permittivity $\overline{\varepsilon}_{zz}=1+4\pi\, d\overline{P_{VZ}}/dE_0\big|_{E_0=0}$ could be obtained from (23)-(25).

In $c_p$, $c$-phase one obtains from (25) that:

$$\begin{cases} \bar{\varepsilon}_{zz}(T,h) = \dfrac{4\pi(1-1/h(1+\Lambda_Z))}{A_m + 3B_m P_{VZ}^2 + 2D_m P_{VZ}} \\ A_m P_{VZ} + D_m P_{VZ}^2 + B_m P_{VZ}^3 + E_m = 0 \end{cases} \qquad (31)$$

In *r*-phase one obtains from (24) that:

$$\begin{cases} \bar{\varepsilon}_{zz}(T,h) = \dfrac{4\pi(1-1/h(1+\Lambda_Z))}{\left(A_m - \dfrac{F_{XZ}A_X - 2K_m^2}{B_X + C_{XY}}\right) + 3\left(B_m - \dfrac{F_{XZ}^2}{B_X + C_{XY}}\right)P_{VZ}^2 + 2\left(D_m + \dfrac{3K_m F_{XZ}}{B_X + C_{XY}}\right)P_{VZ}} \\ \left(A_m - \dfrac{F_{XZ}A_X - 2K_m^2}{B_X + C_{XY}}\right)P_{VZ} + \left(D_m + \dfrac{3K_m F_{XZ}}{B_X + C_{XY}}\right)P_{VZ}^2 + \left(B_m - \dfrac{F_{XZ}^2}{B_X + C_{XY}}\right)P_{VZ}^3 + E_m = 0 \end{cases} \qquad (32)$$

The static dielectric permittivity over film thickness at room temperature calculated with the help of Eq.(31) is presented in Fig.9. At zero external field $E_0 = 0$ the size driven phase transition with permittivity divergence exists only at $U_m = 0$ (see curve 0 and [7]). With $|U_m|$ increase there is a maximum in $\bar{\varepsilon}_{zz}(T,h)$ which shifted to larger *h* values (compare curves 5 and 6). For large enough $|U_m|$ values $\bar{\varepsilon}_{zz}(T,h)$ flattens with thickness increase and have no maxima (see Fig.9) Because curves 1, 2, 3, 4 represent permittivity of PZT film on Pt, SrTiO$_3$, LSAT, LaAlO$_3$ substrates, while curves 5 and 6 with maxima represents the influence of the substrates with several times smaller misfit strain, Fig.9 proves the essential influence of substrates on the film properties. The same conclusion follows from the temperature dependence of $\bar{\varepsilon}_{zz}$ depicted in Fig.10, where curves 1-4 correspond to the aforementioned substrates, while curve 0 for free standing film ($U_m = 0$) with divergence at critical temperature, that corresponds to thickness induced para - ferroelectric phase transition. With $|U_m|$ increase the shift of the maxima to larger temperatures and their broadening take place, the latter being larger with thickness decrease. The tiny finite discontinuous changes related to the phase transition between *c* and *r*-phases are evident for two thinnest films (see $h = 50$ and $h = 100$, they are marked by thin dashed line).

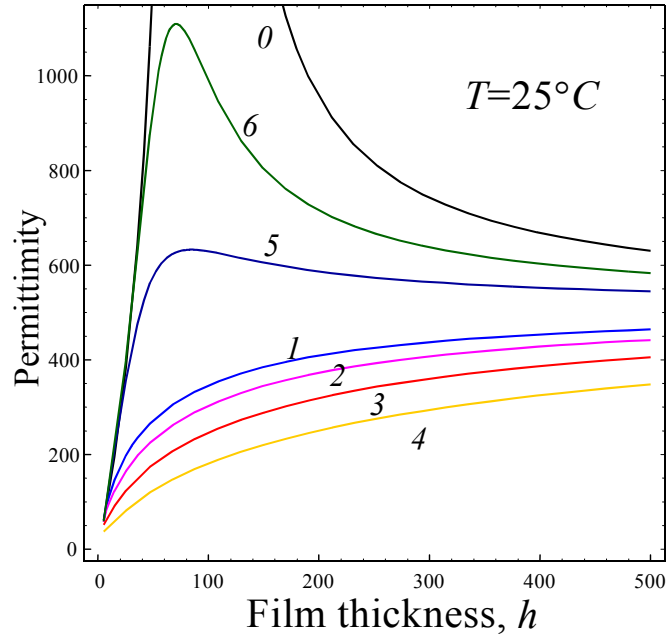

**Figure 9.** Linear dielectric permittivity $\bar{\varepsilon}$ over film thickness $h$ for 50/50 PZT film at $h_d = 10$, $\Lambda = 10$, $R = 10$, $T = 25^0 C$, $E_0 = 0$ and different substrates with $U_m = 0$ (curve 0), $U_m = -0.024$ (Pt, curve 1), $U_m = -0.029$ (SrTiO$_3$, curve 2), $U_m = -0.039$ (LSAT, curve 3), $U_m = -0.06$ (LaAlO$_3$, curve 4) -0.01(curve 5), -0.005 (curve 6).

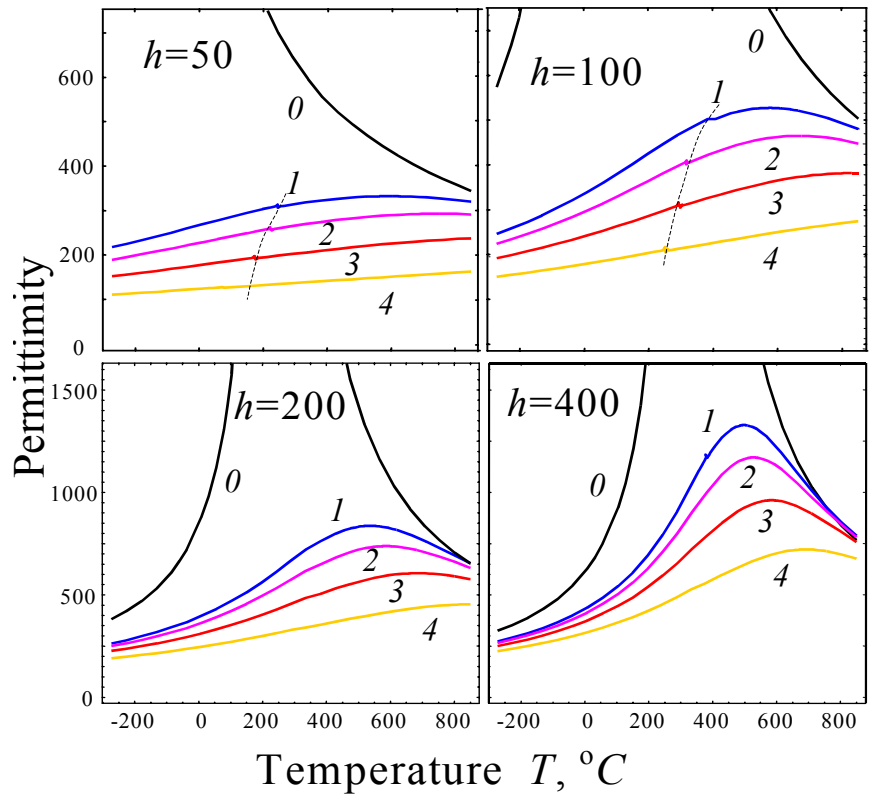

**Figure 10.** Linear dielectric permittivity $\bar{\varepsilon}$ over temperature $T$ for 50/50 PZT film with thickness $h = 100$ at $h_d = 10$, $\Lambda = 10$, $R = 10$, $E_0 = 0$ and different substrates: Pt ($U_m = -0.024$, curve 1), SrTiO$_3$ ($U_m = -0.029$, curve 2), LSAT ($U_m = -0.039$, curve 3), LaAlO$_3$ ($U_m = -0.06$, curve 4).

## DISCUSSION

The mismatch effect considered in this paper is related to mechanical strain tensor $u_{xx}$ originated from the difference in a substrate and a film lattice constants, thermal coefficients and growth imperfections. The mechanical strain influences the electric polarization via electrostriction and via piezoelectric effect on the surface. The latter effect appears even in cubic lattices because of the absence of inversion centre in $z$ direction near the surface.

Our theory predicts that mismatch-induced built-in field $E_m$ and odd power polarization terms in free energy do cause self-polarization in thin ferroelectric films. Necessary conditions (film thickness $h$, misfit strain $U_m$, temperature $T$) are determined by material coefficients and can be calculated for each definite material. As an example we performed calculations for 50/50 PZT films. However some important general features are evident from the performed calculations.

First of all the peculiarities of phase diagram at $h < h_{cr}$, where thickness-induced phase transition is expected for free standing film. In the majority of papers this phase transition was considered as the transition from ferroelectric to paraelectric phase on the basis of calculations with the help of Euler-Lagrange equations for the case $P_X = P_Y = 0$, $P_Z \neq 0$. The solutions of three Euler-Lagrange equations for nonzero three components of polarization have shown that the sense of thickness induced phase transitions is disappearance of $P_Z$ component only at critical thickness or temperature, so that in-plane polarization conserves.

## APPENDIX A

In accordance with (12a) at $h \gg 1$ one obtain, that:

$$\overline{\varphi(z)} = \frac{1}{l}\int_{-l/2}^{l/2} dz \frac{ch(z/l_Z)}{ch(l/2l_Z)+(\lambda_Z/l_Z)sh(l/2l_Z)} = \frac{2l_Z}{l} \cdot \frac{sh(l/2l_Z)}{ch(l/2l_Z)+(\lambda_Z/l_Z)sh(l/2l_Z)} \approx \frac{1}{h(1+\Lambda_Z)} \quad (A.1)$$

At $\alpha_X > 0$ we obtain from (12b) that

$$\overline{\phi(z)} = \frac{1}{l}\int_{-l/2}^{l/2} dz \frac{ch(z/l_X)}{ch(l/2l_X)+(\lambda_X/l_X)sh(l/2l_X)} = \frac{2l_X}{l} \cdot \frac{sh(l/2l_X)}{ch(l/2l_X)+(\lambda_X/l_X)sh(l/2l_X)}. \quad (A.2a)$$

At $\alpha_X < 0$ and $l < \pi l_X$ we obtain from (11b) that

$$\overline{\phi(z)} = \frac{1}{l}\int_{-l/2}^{l/2} dz \frac{\cos(z/l_X)}{\cos(l/2l_X)+(\lambda_X/l_X)\sin(l/2l_X)} = \frac{2l_X}{l} \cdot \frac{\sin(l/2l_X)}{\cos(l/2l_X)+(\lambda_X/l_X)\sin(l/2l_X)} \quad (A.2b)$$

At $l < l_X$ one obtains that $\cos(l/2l_X) \approx ch(l/2l_X) \approx 1$, $\sin(l/2l_X) \approx sh(l/2l_X) \approx l/2l_X$, and so

$$\overline{\phi(z)} \approx \frac{2l_X}{l} \cdot \frac{(l/2l_X)}{1+(\lambda_X/l_X)(l/2l_X)} = \frac{1}{1+\lambda_X l/2l_X^2},$$ 

independently of $\alpha_X$ sign. Thus we calculate that

$1 - \overline{\phi(z)} \approx 1 - \dfrac{1}{1 + \lambda_X l / 2 l_X^2} = \dfrac{1}{1 + 2 l_X^2 / \lambda_X l} \equiv \dfrac{1}{1 + R/h}$. Approximate expressions for higher coefficients could be obtained similarly, but much more cumbersome.